\documentclass[pdflatex,sn-mathphys-num]{sn-jnl}

\usepackage{graphicx}%
\usepackage{multirow}%
\usepackage{amsmath,amssymb,amsfonts}%
\usepackage{amsthm}%
\usepackage{mathrsfs}%
\usepackage[title]{appendix}%
\usepackage{xcolor}%
\usepackage{textcomp}%
\usepackage{manyfoot}%
\usepackage{booktabs}%
\usepackage{algorithm}%
\usepackage{algorithmicx}%
\usepackage{algpseudocode}%
\usepackage{listings}%

\theoremstyle{thmstyleone}%
\theoremstyle{thmstyletwo}%

\theoremstyle{thmstylethree}%

\begin{document}

\title[Circular transformation of steel industry.]{Circular transformation of the European steel industry renders scrap metal a strategic resource}


\author*[1,2,3,4]{\fnm{Peter} \sur{Klimek}}\email{peter.klimek@ascii.ac.at}

\author[1]{\fnm{Maximilian} \sur{Hess}}\email{maximilian.hess@ascii.ac.at}

\author[1,5]{\fnm{Markus} \sur{Gerschberger}}\email{markus.gerschberger@ascii.ac.at}

\author[1,2,3,6]{\fnm{Stefan} \sur{Thurner}}\email{thurner@csh.ac.at}

\affil*[1]{\orgname{Supply Chain Intelligence Institute Austria}, \orgaddress{\street{Josefstädter Strasse 39}, \city{Vienna}, \postcode{1080}, \country{Austria}}}

\affil[2]{\orgdiv{Institute of the Science of Complex Systems, CeDAS}, \orgname{Medical University of Vienna}, \orgaddress{\street{Spitalgasse 23}, \city{Vienna}, \postcode{1090},\country{Austria}}}

\affil[3]{\orgname{Complexity Science Hub}, \orgaddress{\street{Josefstädter Strasse 39}, \city{Vienna}, \postcode{1080}, \country{Austria}}}

\affil[4]{\orgdiv{Division of Insurance Medicine, Department of Clinical Neuroscience}, \orgname{Karolinska Institutet}, \orgaddress{\city{Stockholm}, \postcode{17177},\country{Sweden}}}

\affil[5]{\orgdiv{Josef Ressel Centre for Real-Time Value Network Visibility, Logistikum},\orgname{University of Applied Sciences Upper Austria}, \orgaddress{\street{Wehrgrabengasse 1-3}, \city{Steyr}, \postcode{4400}, \country{Austria}}}

\affil[6]{\orgname{Santa Fe Institute}, \orgaddress{\street{1399 Hyde Park Road}, \city{Santa Fe}, \postcode{87501}, \state{NM}, \country{USA}}}

\abstract{The steel industry is a major contributor to CO$_2$ emissions, accounting for 7\% of global emissions. The European steel industry is seeking to reduce its emissions by increasing the use of electric arc furnaces (EAFs), which can produce steel from scrap, marking a major shift towards a circular steel economy. Here, we show by combining trade with business intelligence data that this shift requires a deep restructuring of the global and European scrap trade, as well as a substantial scaling of the underlying business ecosystem. We find that the scrap imports of European countries with major EAF installations have steadily decreased since 2007 while globally scrap trade started to increase recently. Our statistical modelling shows that every 1,000 tonnes of EAF capacity installed is associated with an increase in annual imports of 550 tonnes and a decrease in annual exports of 1,000 tonnes of scrap, suggesting increased competition for scrap metal as countries ramp up their EAF capacity. Furthermore, each scrap company enables an increase of around 79,000 tonnes of EAF-based steel production per year in the EU. Taking these relations as causal and extrapolating to the currently planned EAF capacity, we find that an additional 730 (SD 140) companies might be required, employing about 35,000 people (IQR 29,000-50,000) and generating an additional estimated turnover of USD 35 billion (IQR 27-48). Our results thus suggest that scrap metal is likely to become a strategic resource. They highlight the need for a massive restructuring of the industry's supply networks and identify the resulting growth opportunities for companies.}




\maketitle

\section{Introduction}\label{sec1}

The steel industry is a major contributor to CO$_2$ emissions. The industry accounts for 7\% of total CO$_2$ emissions worldwide and 5\% in Europe \cite{somers2022technologies}. In countries with a strong steel industry, these shares could be even higher. For example, in 2021 the steel industry accounted for about 16\% of Austria's emissions, or about a third of Austria's total industrial emissions \cite{voestcr}. Most of these emissions, 12.4 Mt in total, come from one company ranking among the European industrial facilities with the highest global warming potentials \cite{erhart2023environmental}. Overall, these rankings are dominated by industrial facilities engaged in electricity production and iron- and steelmaking \cite{erhart2023environmental}.

The European steel industry has therefore embarked on a path to reduce its emissions by 80--95\% by 2050 compared to 1990 \cite{somers2022technologies}. This will be achieved through a combination of low carbon projects, including direct carbon abatement, carbon smart use and carbon capture and storage activities \cite{EuroferProjects}. 

Key to these low CO$_2$ projects is the increased use of electric arc furnaces (EAFs) \cite{vogl2018assessment}. In these furnaces, a mixture of scrap metal and direct reduced iron (sponge iron) is brought into contact with an electric arc and melted. By 2023, 62\% of the world's steelmaking capacity uses the more carbon-intensive  basic oxigen furnace (BOF) production route, while 29\% uses EAFs. Current planned steelmaking capacity is 43\% EAF, but this is still short of the estimated 53\% EAF steelmaking capacity needed in 2050 to meet emissions targets \cite{swalec2023pedal}. Therefore, further efforts are needed to meet these targets.

These ambitious targets require a massive restructuring of existing production processes and structures, which poses significant challenges \cite{pauliuk2013steel, kim2022critical}. One well known challenge is the significant energy consumption of EAFs \cite{manana2021increase}. The energy cost of EAFs can be more than 20\% of the total cost. Several recent studies have therefore addressed the electrical and chemical energy consumption of EAFs using statistical and predictive modelling approaches to optimise their consumption with respect to various operational aspects \cite{carlsson2020modeling,tomavzivc2022data}.  

In addition to the operational challenges at the level of individual installations, the increased capacity of EAFs also puts a strain on the electricity transmission infrastructure. EAFs have been shown to be a potential source of significant voltage fluctuations in electricity grids, leading to flickering of lights \cite{lukasik2020estimating} and suggesting the need for demand-side management strategies in the absence of new transmission infrastructure \cite{manana2021increase}. However, the transition of the European steel industry to zero emissions by 2050 is expected to require 400 TWh of CO$_2$-free electricity, which represents a sevenfold increase over the industry's current consumption or 13\% of the EU's current total electricity production \cite{roadmap2019pathways}.  

Low-cost renewable electricity is therefore a key enabler of the steel industry's green transition. Statistical modelling from a systems perspective for Northern Europe suggests that this increased demand can be met by wind and solar power in these regions, providing a cost-effective transformation pathway \cite{toktarova2022interaction}. The extent to which this is feasible elsewhere is unclear. On a global scale, it has been estimated that reducing CO$_2$ emissions from the iron and steel industry by almost a third would require an investment of USD 0.9 trillion, and the energy required would be equivalent to 1\% of global supply \cite{gielen2020renewables}.  It has also been suggested that green hydrogen-based steelmaking facilities should be co-located with iron ore deposits and renewable energy opportunities (solar and wind power) rather then with existing facilities to meet these challenges \cite{devlin2023global}. 

Here we show that the required restructuring of production processes and structures triggered by a shift to EAFs goes far beyond the demand for low-emission electricity. Rather, the widespread adoption of EAFs requires a steady stream of scrap metal as an input, and thus an entire ecosystem of companies involved in the recycling and transport of scrap metal. Indeed, the international scrap trade forms a complex and dynamically changing network in which individual countries play highly heterogeneous roles, including the emergence of trading communities and hierarchical relationships that reflect geopolitical and economic turbulence \cite{hu2020characteristics}. 

We combine comprehensive trade data and business intelligence data on individual companies to show that the European scrap trade and the population of companies with related activities strongly correlate with existing EAF capacity in individual European countries. By assuming that these observed correlations are causal and extrapolating these relationships to currently planned EAF capacity \cite{EuroferProjects}, we quantify the changes required in the scrap trade and underlying companies.

\section{Data and Methods}\label{secm}

{\bf Scrap metal trade networks.} Data on bilateral trade flows for about 200 countries going back to 2007 are used, as provided by the BACI dataset \cite{CEPII:2010-23}. Trade flows in ferrous scrap are extracted as flows associated with the Harmonised System code 7204 (ferrous waste and scrap; remelting scrap ingots of iron or steel) expressed in quantities. The data are grouped into three time intervals, 2007-2011, 2012-2016 and 2017-2021, and the values for each bilateral trade flow are averaged over the respective time windows. Imports and exports for each country are calculated as the sum of all incoming and outgoing flows. For visualisation purposes, we show the backbones of the resulting networks after removing non-significant links using the disparity filter \cite{serrano2009extracting}.

{\bf Business intelligence data.}  Commercial company data from Bureau van Dijk is used, the Orbis database \cite{van2023private}. It contains information on several hundred million companies worldwide. Despite this comprehensiveness, care must be taken when extracting information from Orbis as its coverage and representativeness may vary across countries. In the present analysis, we use a snapshot (bulk downloaded) version of the data from 2023 and focus only on European countries in the statistical modelling, where coverage of the company population is typically considered to be higher than for some other world regions \cite{/content/paper/c7bdaa03-en}. Furthermore, we focus only on companies for which Orbis provides comprehensive information in the form of a textual description of their business activities (i.e. non-empty fields in at least one of the columns "Full Overview", "Main Products and Services", "Main Activity" and "Primary Business Line"). Within these fields, we extract all companies that contain the keyword "scrap" and extract them along with their country, primary industry classification (NAICS), operating revenue and number of employees.

{\bf Topic modelling.} For a more accurate description of company activities, topic modelling using Latent Dirichlet Allocation (LDA) is performed on the textual description of company activities \cite{foulds2013stochastic}. In this approach, each business description ("document") is described as a mixture of topics. Each topic is in turn described by a vector of word frequencies. LDA is a widely used natural language processing technique \cite{blei2003latent}; we used the implementation provided in MatLab R2021a.

The text was pre-processed using standard MatLab procedures (see its documentation). The text was tokenised and lemmatised; punctuation was removed. Words of three characters or less were then removed, along with stop words, for which we used a list of the 1,000 most frequent words as extracted from a corpus of New York Times articles \cite{dodds2011temporal}, enriched with non-specific words often used in the Orbis company descriptions.

LDA requires as input the number of topics to be extracted from the documents. To this end, we used 10\% of the documents as hold out data and fitted varying numbers of topics to the remaining training data. The goodness of fit of the resulting LDA model on the held-out data was then assessed using the perplexity \cite{wallach2009evaluation}, and the number of topics giving minimal perplexity was used in the remainder of the analysis.

{\bf Statistical modelling.} We represent the distribution of numerical firm variables (number of employees, operating revenue) as cumulative distribution functions, $CDF(x)$, defined as the probability of observing a value of a random variable $x$ that is less than or equal to $CDF(x)$. The inverse transformation method is used to generate estimates for the firm variables (employees, operating revenue) that are compatible with the observed empirical distribution. 1,000 randomly sampled firm populations are generated for each country. The variables are summed over all firms in each iteration and the median and interquartile range (IQR) are computed as the percentiles of the distribution of theses sums over all iterations.

EAF capacity is related to scrap trade data, currently installed capacity of BOFs (which can also utilize scrap metal) and company information by means of a linear regression model with fixed effects and no intercepts. In this model, the dependent variable is the EAF capacity (planned or currently installed) per country. The independent variables are the country's scrap imports and exports, the currently installed BOF capacity \cite{swalec2023pedal}, the number of scrap-related enterprises and their combined turnover and number of employees.

\section{Results}\label{sec2}

\subsection{Evolution of the global scrap metal trade network}\label{subsec2}

\begin{figure}[h]
\centering
\includegraphics[width=0.9\textwidth]{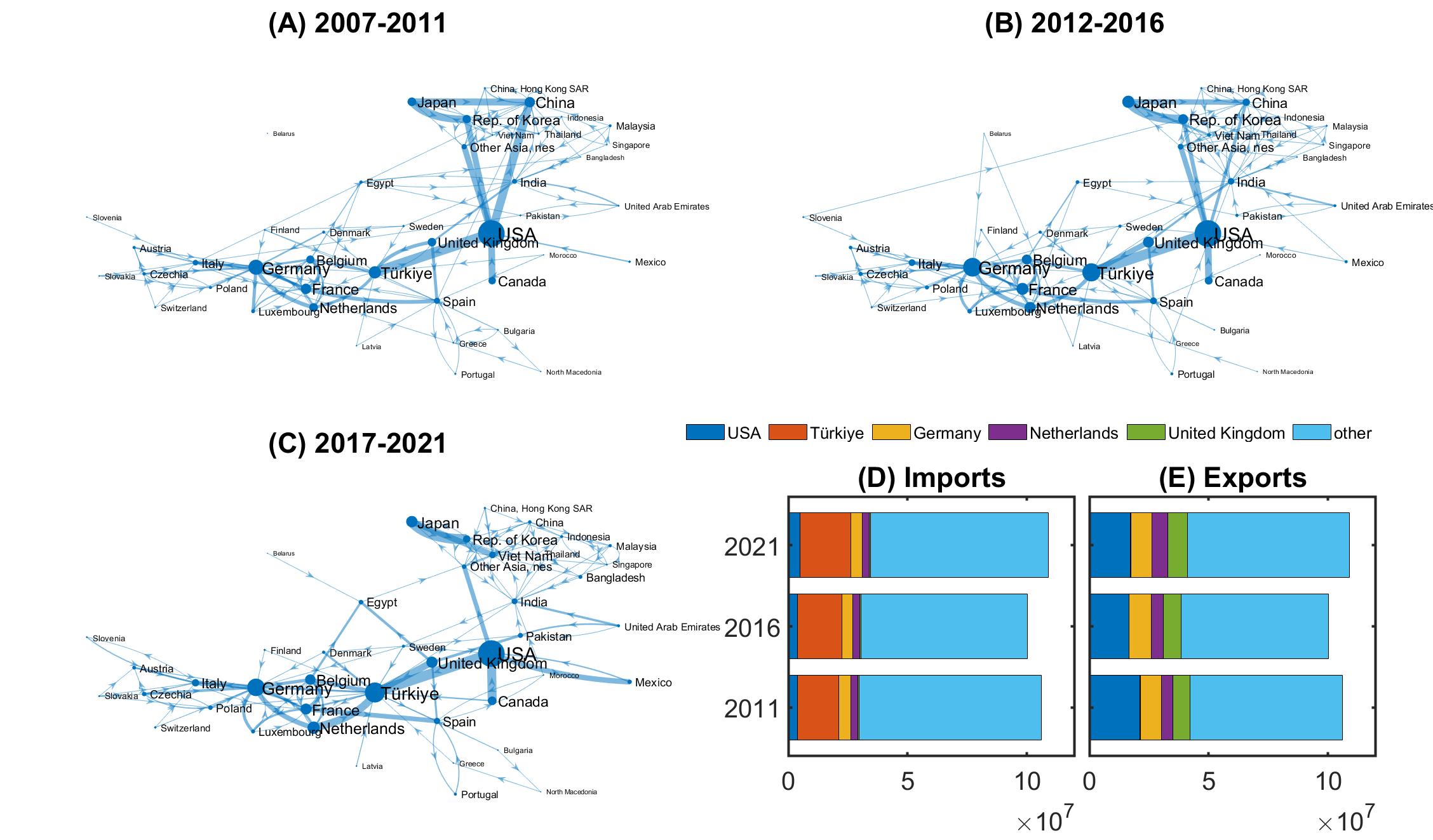}
\caption{Evolution of the global scrap metal trade network from (A) 2007-2011 over (B) 2012-2016 to (C) 2017-2021. Directed links show trade flows with a proportional thickness. Node sizes are proportional to the combined imports and exports. Note how China decouples from the rest of the world over time despite being the world's largest steel producer. Overall, both (D) imports and (E) exports decrease over from 2011 to 2016 and then increase again until 2021.}\label{fig:fig1}
\end{figure}

In Figure~\ref{fig:fig1} we show the evolution of the global scrap metal trade network over the last 15 years. We show three time periods comprising the average annual bilateral trade in (A) 2007-2011, (B) 2012-2016 and (C) 2017-2021. Looking at the combined values of all (D) imports and (E) exports, we see a trend of declining trade in scrap from the first to the second period, followed by an increase in global trade from the second to the third.

A central country in the trade network is the US. Initially, in 2007-2011, the US had strong links with several Asian countries, including China and the Republic of Korea. However, more recent trade data show that these links have been severed. The main trading partners of the US are now Mexico, Canada and Turkey. Turkey plays a special role in the global trade network. It is the world's largest importer of ferrous scrap, but its exports are orders of magnitude smaller than its imports; it is a sink in the trade network. 

China's exports and imports of ferrous scrap have both fallen dramatically over time, while Vietnam and Japan have become more important over time. European countries, on the other hand, show remarkably stable trade relations over the period.

\begin{figure}[h]
\centering
\includegraphics[width=0.9\textwidth]{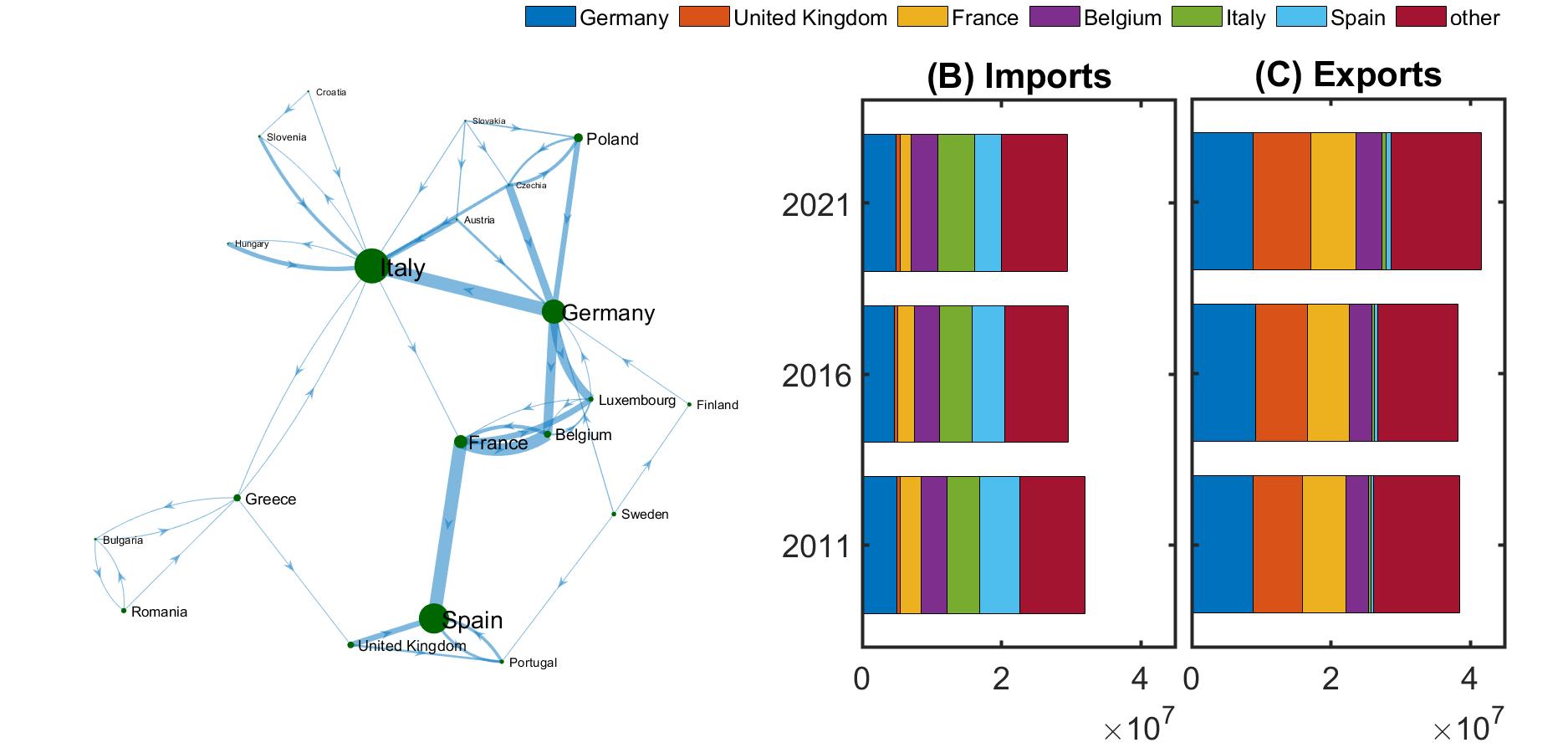}
\caption{Scrap metal trade network between all European countries with major EAF installations (A), along with the overall (B) imports and (C) exports. Total imports decrease over time while exports started to increase since 2016. Germany and France are net exporters while Italy and Spain are net importers, in line with their high installed EAF capacity.}\label{fig:fig2}
\end{figure}

Figure~\ref{fig:fig2}(A) gives a more detailed view of the European scrap trade network of countries with major EAF installations. Note that the Netherlands is not part of this network as it plays an important role in shipping logistics but not yet in EAF steelmaking. However, with major new installations planned in the Netherlands, this is likely to change in the future.

Figure~\ref{fig:fig2}(B,C) shows the corresponding imported and exported ferrous waste over time. Overall, combined imports show a decreasing trend over time.
This decrease also persists into the time period from 2017-2021 in contrast to global scrap trade, which expanded in this time period.
Germany and France are net exporters while Italy and Spain, both of which have significant EAF capacity, are net importers.
Exports are considerably larger than imports in all time periods and have increased from 2016 to 2021, in contrast to the imports.
Hence, these European countries are net exporters of scrap metal with a strongly decreasing import-to-export ratio in recent years.

The size of the nodes in Figure~\ref{fig:fig2} is proportional to the current installed EAF capacity of the countries. The highest capacities are found in Italy, Spain, Germany and France. These countries are also the destination of the main EU scrap trade flows.

\begin{figure}[h]
\centering
\includegraphics[width=0.9\textwidth]{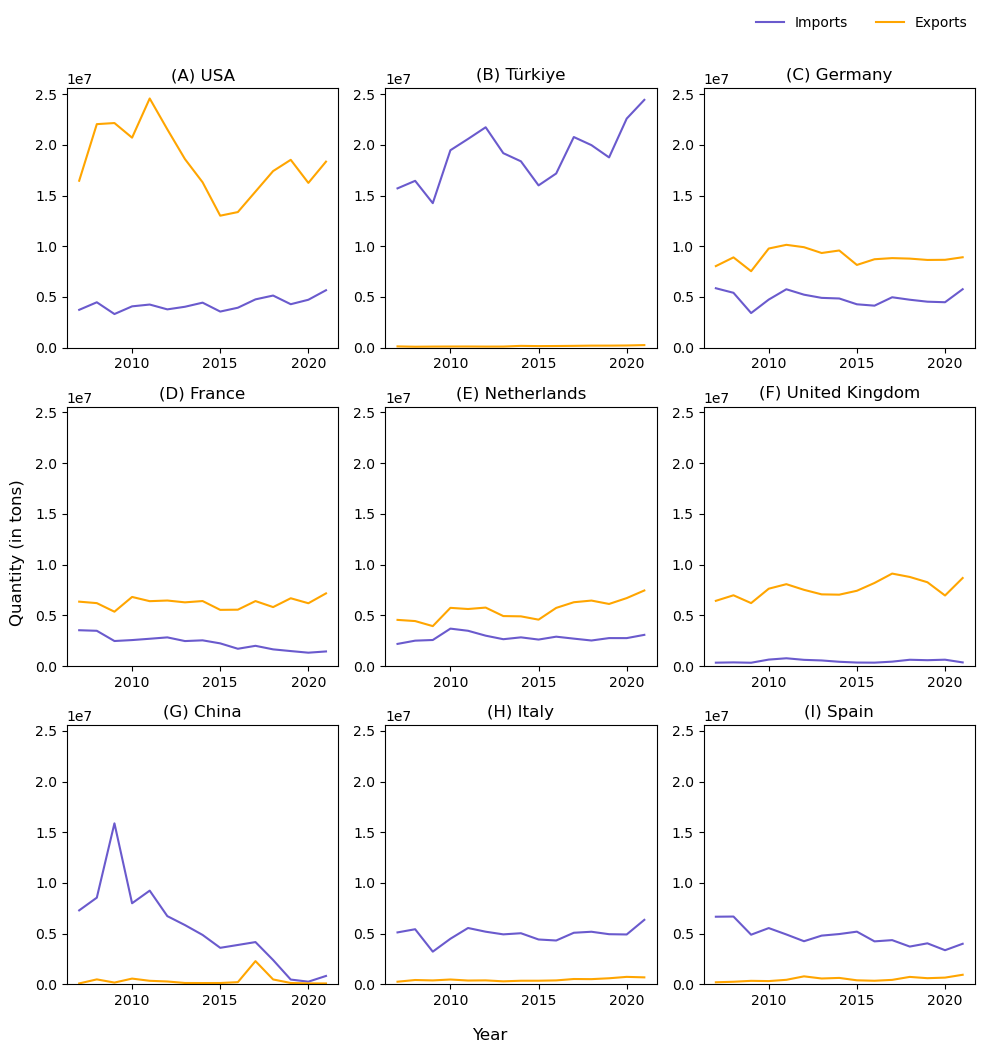}
\caption{Scrap imports and exports by country. Results for imports (blue) and exports (orange) are shown for the US, Turkey, China and the top six European traders from 2007 to 2021. There is a great deal of variability between countries in terms of their import/export ratios and whether these flows are increasing or decreasing over time.}\label{fig:countries}
\end{figure}

Figure~\ref{fig:countries} shows that the United States (A), Germany (C), France (D), the Netherlands (E), and the United Kingdom (F) are net exporters of scrap, meaning that their exports are consistently higher than their imports. Turkey (B), China (G), Italy (H) and Spain (I) are net importers, with exports considerably lower than their imports. In recent years, after 2017, both Chinese imports and exports are close to zero. Turkey has an increasing trend in imports, in contrast to France and Spain. Most other countries do not show a consistent rising or falling trend in their trade volume.  

\subsection{Characterizing the scrap metal firm ecosystem}\label{subsec3}

We identified 5,227 companies with scrap-related business activities. These companies employ approximately 390,000 people and have a combined operating income of USD 420 billion. 46.8\% report NAICS code 4239 (miscellaneous durable goods wholesalers) as their primary activity, 18.5\% are primarily metal and mineral (except petroleum) wholesalers (NAICS 4235), 7.7\% report remediation and other waste management services as their primary activity (NAICS 5629), and 3.3\% are chemical and allied products wholesalers (NAICS 4246). All other activities were carried out by less than 3\% of enterprises. 

\begin{figure}[h]
\centering
\includegraphics[width=0.9\textwidth]{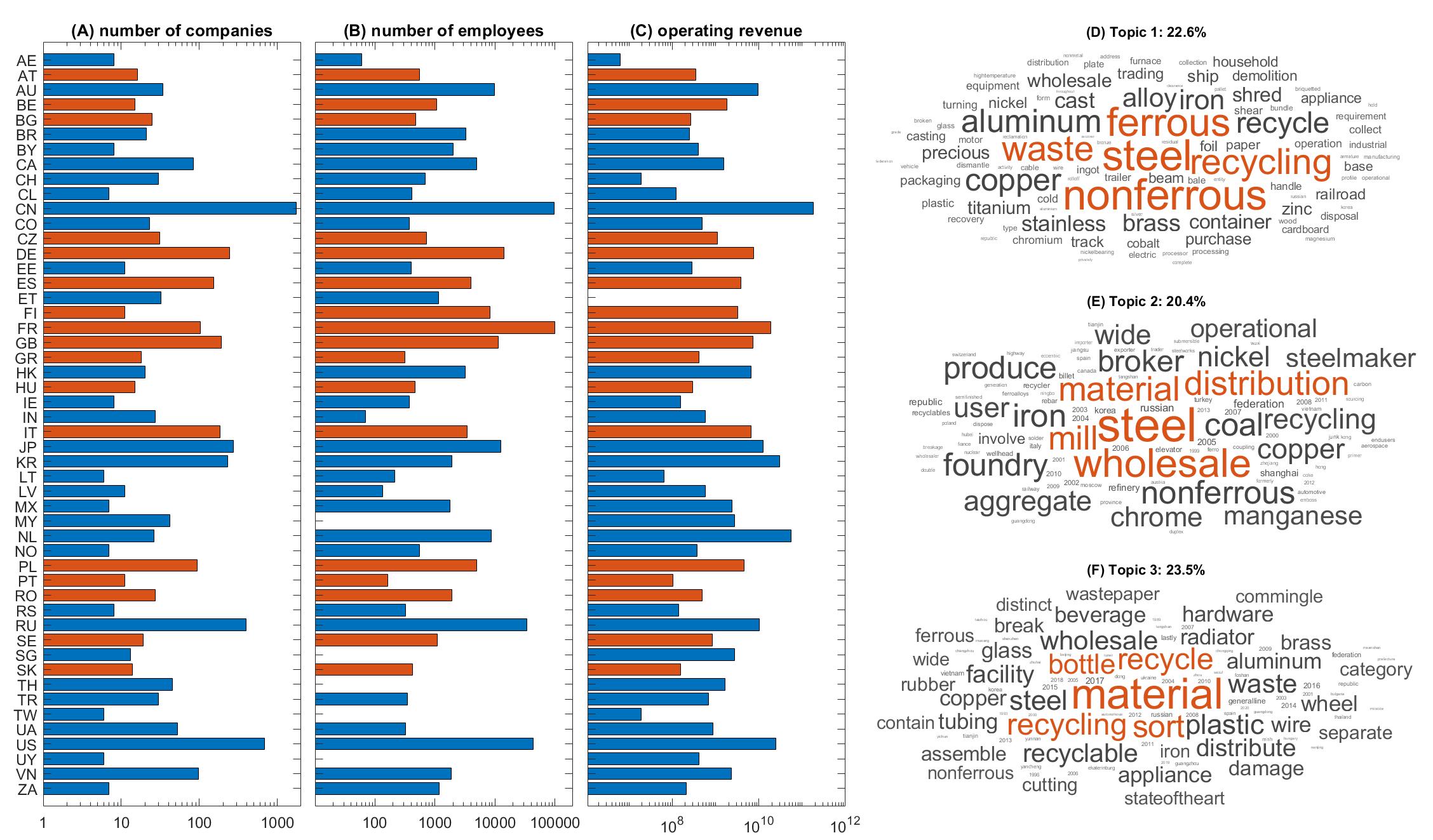}
\caption{Description of scrap-metal-related companies. We show (A) the number of companies related to scrap metal and their combined (B) number of employees and (C) operating revenue with the main EU countries with EAF installations being highlighted in red. The three main topics describing the activities of these companies as derived from the LDA analysis are shown as word clouds (D-F), in which the word size is proportional to the relevance of the term to the corresponding topic.}\label{fig:fig3}
\end{figure}

Figure~\ref{fig:fig3} gives an overview of the distribution of these enterprises across countries (for all countries with more than 5 enterprises) with (A) the number of enterprises per country and their combined (B) number of employees and (C) operating turnover. European countries with significant EAF capacity are shown in red (note the logarithmic x-axes). Globally, most scrap-related companies are concentrated in China, the US, Russia, Japan, Korea, Italy, the UK, Germany and Spain. In terms of employees and turnover, this shifts subtly, with France, for example, playing a more prominent role than its number of companies would suggest. 

Another outlier is the Netherlands. They have one of the highest operating revenues, but don't currently have any significant EAF capacity. This is likely related to the central role of Dutch ports in the transport sector.

In order to gain a better understanding of the firm's activities, we perform a topic modelling analysis of the textual firm descriptions. We find that most of these texts can be characterised by a mixture of three topics, each contributing more than 20\%; all other topics contribute less than 10\%, see Figure~\ref{fig:fig3}(D)-(F). 

The identified topics mirror the main activities in metal recycling. In brief, the value chain of such companies typically contains the following three steps. Companies buy batches of materials from steel manufacturers or foundries (to which several important terms from topic 2 refer) or from other companies collecting, shipping and trading scrap metals from demolitions sites (several medium frequency terms in topic 1). In the next steps, companies sort and recycle various types of waste (main terms from topic 3), before selling the ferrous and nonferrous fractions that have been recovered (topic 1). Therein, scrap parts are shredded to retrieve the ferrous fractions. This secondary raw materials that can again be used by steelmakers in EAFs. Hence, topic 2 loosely relates to the first step of the value chain (scrap input), topic 3 to the second step of sorting by category and the selling of the separated ferrous and nonferrous fractions can be found in topic 1.

\begin{figure}[h]
\centering
\includegraphics[width=0.9\textwidth]{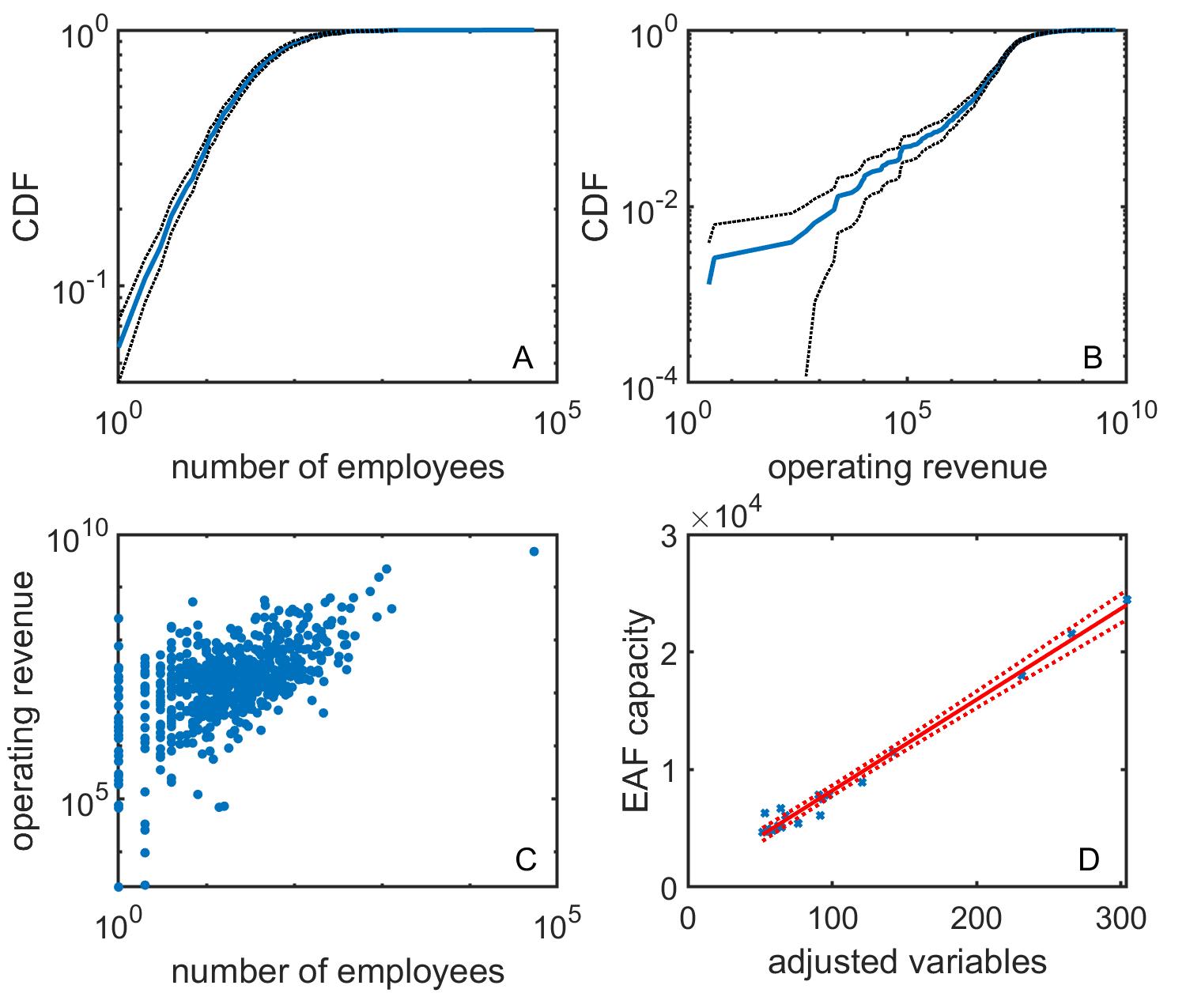}
\caption{Statistical modelling of scrap-metal-related companies in the EU. The cumulative distribution functions (CDF) of (A) their number of employees and (B) operating revenue are shown on a double logarithmic scale along with their 95\% confidence bounds, revealing a power-law-like scaling for the operating revenue. A scatter plot of these variables shows their correlation with each other (C). Including these variables in a regression model (red line with 95\% confidence bounds) along with countries' imports and exports provides an excellent fit of the installed EAF capacity (D). }\label{fig:fig4}
\end{figure}

Figure~\ref{fig:fig4} shows more detailed statistical characteristics of EU scrap related enterprises. The empirical cumulative distribution function (CDF) of the number of employees and operating revenues for these companies is shown in Figure~\ref{fig:fig4}(A) and (B) respectively. It can be seen that the distribution of operating revenue in particular scales over several orders of magnitude across the enterprises, in contrast to the CDF of the number of persons employed, which tends to cluster around values of 100 persons employed. Nevertheless, operating turnover and number of employees tend to be correlated, as can be seen in Figure~\ref{fig:fig4}(C) and from their Pearson correlation coefficient of 0.55 ($p<0.0001$).

\subsection{EAF capacity strongly correlates with European scrap metal trade}\label{subsec4}

To explore the relationships between scrap trade, company activities and EAF capacity in more detail, a regression analysis is carried out, where each country corresponds to one observation and its current EAF capacity is the dependent variable. The independent variables include the country's combined exports and imports in 2017-2021, the currently installed BOF capacity, the number of scrap metal-related companies, their number of employees and their combined turnover, see Figure~\ref{fig:fig4}(D).

\begin{table}[h]
\caption{Regression model linking scrap metal trade and company activities with EAF capacity. The table gives the coefficient estimate along with its standard deviation (SD) and p-value for the fixed effect regression. Significantly correlated coefficients ($p<0.01$) are highlighted in boldface.}\label{tab1}%
\begin{tabular}{@{}lll@{}}
\toprule
Variable & Estimate (SD)  & p-Value \\
\midrule
Exports 2017--2021    & {\bf $-0.00096(14)$} & {\bf $<0.0001$}  \\
Imports 2017--2021    & {\bf $0.0018(2)$} &  {\bf $<0.0001$}  \\
Number of companies   & {\bf $79(11)$}   &  {\bf $<0.0001$}  \\
Number of employees	& $0.13(8)$ & $0.11$ \\
Operating revenue & $-2.4(3.7) \cdot  10^{-7}$ & $0.53$ \\
BOF capacity & $-0.12(6)$  & $0.08$ \\
\botrule
\end{tabular}
\end{table}

The model provides a very good fit to the data with an adjusted $R^2$ of $0.98$, meaning that a country's trade flows and population of scrap-related enterprises explain most of the variation between their EAF capacities. Higher EAF capacity is associated with increased imports, decreased exports and an increase in the number of scrap-related enterprises, see Table~\ref{tab1}. 
The number of employees and operating turnover are not significantly correlated with EAF capacity, as is BOF capacity.

More specifically, if the observed correlations were causal, these results suggest that each additional scrap company enables an increase in EAF-based steelmaking of around 79,000 tonnes per year. Furthermore, an increase in annual imports of 550 tonnes and a decrease in annual exports of 1,000 tonnes coincide with an increase in EAF capacity of 1000 tonnes.

\subsection{Towards an estimation of the scrap metal company ecosystem for a CO$_2$ free steel industry}\label{subsec5}

We use the strong phenomenological correlation between EAF capacity, trade flows and the business ecosystem to extrapolate these variables to their expected values implied by currently planned EAF capacities. We use the correlation between EAF capacity and the number of scrap-related companies from Table~\ref{tab1} to scale the number of companies in a country to the level implied by the planned EAF capacity. This implicitly assumes that this planned  EAF installations will be operated with the same capacity and productivity as the currently installed EAF facilities. 

Note that operating revenue and the number of employees have a much weaker and non-significant correlation with EAF capacity than the number of firms. Therefore, in order to estimate the additional revenue and employees of the additional scrap companies, we sample these values for each country from their empirical distribution function. 

Table~\ref{tab2} shows the results for the additional number of companies, their turnover and employees as implied by the planned EAF capacity for each country. Overall, extrapolating from the current EAF capacity suggests that the planned EAF capacity implies an additional 730 (SD 140) companies with an estimated (median) number of 35,000 employees (IQR 29,000-50,000) and an estimated turnover of USD 35 billion (IQR 27-48).

\begin{table}[h]
\caption{Estimating the scrap-metal-related company ecosystem for the planned EAF capacity of countries. The number of additional companies (along with its SD) implied by the planned EAF capacity is given. Operating revenue and the number of additional employees are sampled for these additional companies from the corresponding CDFs and reported by their median and inter-quartile range (IQR).}\label{tab2}
\begin{tabular*}{\textwidth}{@{\extracolsep\fill}l p{2cm} p{2cm} p{2cm} p{2cm}}
\toprule%
Country & Planned EAF Capacity	& Additional Companies (SD)	& Additional Revenue (IQR) [bn USD] & Additional Employees (IQR) [thds.] \\
\midrule
Austria	& 2450	& 31(6)	& 1.2 (1.0-1.8)	& 1.3 (1.1-1.9) \\
Belgium	& 2500	& 32(6)	& 1.3 (1.1-1.7)	& 1.4 (1.1-1.8) \\
Croatia	& 200	& 3(0.5)	& 0.07 (0.03-0.14)	& 0.08 (0.04-0.16) \\
Czechia	& 3500	& 44(8)	& 1.9 (1.4-2.7)	& 2.0 (1.5-2.8) \\
Finland	& 5100	& 65(12)	& 3.0 (2.3-3.9)	& 3.1 (2.5-3.7) \\
France	& 6500	& 82(16)	& 3.8 (3.3-5.5)	& 4.0 (3.4-5.0) \\
Germany	& 17600	& 223(42)	& 12 (10-15)	& 12 (10-19) \\
Italy	& 2500	& 32(6)	& 1.4 (0.98-2.7)	& 1.1 (1.0-2.5) \\
Luxembourg 	& 250	& 3(1) &	0.05 (0.03-0.10)	& 0.07 (0.05-0.16) \\
Poland	& 1000	& 13(2)	& 0.54 (0.33-0.98)	& 0.57 (0.37-0.91) \\
Romania	& 4100	& 52(10)	& 2.3 (1.9-3.2)	& 2.5 (2.0-3.2) \\
Spain	& 1700	& 22(4)	& 0.86 (0.70-1.2)	& 0.90 (0.75-1.3) \\
Sweden	& 9200	& 117(22)	& 6.2 (4.7-8.1)	& 6.2 (5.0-7.3) \\
United Kingdom	& 780	& 10(2)	& 0.38 (2.4-5.6)	& 0.41 (0.27-0.59) \\
\botrule
\end{tabular*}
\end{table}

\section{Discussion}\label{sec12}

In this paper, we quantitatively assess how the greening of the EU steel industry requires a massive restructuring of the industry's supply networks. We combine information on currently planned and installed EAF capacity across EU countries with scrap trade and business intelligence data to show that the population of scrap-related firms and their associated trade flows closely track EAF capacity. We use the phenomenological relationships between these variables to estimate how the business and trade system can be expected to scale as EAF capacity increases.

Considering trade data over 15 years from 2007 to 2021, we find a heterogeneous dynamics of global scrap metal trade flows. We observe a decoupling of Chinese scrap metal trade flows from the rest of the world. Our observation period coincides indeed with a strong increase in Chinese steel production \cite{ren2021review}, suggesting that China achieves this growth autonomously in terms of scrap. On a European level we observe a contraction of imports, driven by decreasing imports in France and Spain, in line with a reduction of EU steel production, while exports started to increase after 2016. We find that Germany and France are net exporters whereas Italy and Spain, both of which boasting substantial EAF capacity, are net importers of scrap metal. Overall, European countries with major EAF installations are net exporters of scrap with a decreasing import-to-export ratio.

These observations of the trade network dynamic, coupled with the planned substantial increase in EAF capacity, are worrying. The average lifespan of steel products is 40 years, for buildings and infrastructure it can be up to a hundred years \cite{worldsteel}. Global steel demand is greater than the amount of scrap available for recycling and this is unlikely to change in the near future. It is therefore unrealistic to expect global steel demand to be met entirely by recycling. The shift towards a sustainable and circular steel industry makes scrap a new strategic resource. The emergence of this competition for scrap can already be seen in the shifts in trade networks, with China's decoupling from the rest of the world and the negative association between scrap exports and EAF capacity observed in our regression analysis. The  volume of planned EAF projects suggests that competition for scrap is bound to intensify.

Lists of critical and strategic commodities have traditionally focused on energy resources (oil, natural gas, coal), critical minerals and metals, food, water, advanced technologies related to telecommunications and cybersecurity, and defense equipment and weapons. Our findings suggest that scrap will become increasingly critical to manufacturing while also being a key enabler of a more sustainable and circular production system, making it a strategic resource. 

Widespread adoption of EAF technology will require new scrap supply networks. We have mapped scrap-related businesses around the world and identified more than 5,000 companies with a related focus. A topic analysis shows that these companies are indeed focused on the wholesale and distribution of ferrous and non-ferrous scrap and the recycling of various materials.  We then used a regression model to extrapolate imports, exports, and the number of scrap-related companies to planned EAF capacity in Europe, taking advantage of the observed strong correlations (adjusted R-squared of 0.98) between these variables. In doing so, we make several assumptions, namely (1) that the planned EAF capacity will be realized, (2) that there will be sufficient steel demand to operate these facilities at similar capacity levels as today, and (3) that there is indeed a causal relationship in which EAF capacity drives the number of scrap companies, i.e., that most of the business of these companies actually consists of transporting scrap to EAF facilities. Using these assumptions, we estimate that an additional 730 (SD 140)  enterprises are needed in Europe with an estimated (median) number of 35,000 employees (IQR 29,000-50,000) and an estimated turnover of USD 35 billion (IQR 27-48). Our results therefore highlight the opportunities of the green transition, as a new scrap market is emerging.

Given the uncertainties surrounding several supply-side factors (sufficient supply of scrap and renewable energy required for green steelmaking) and demand-side factors (i.e., global demand for steel), the linear extrapolations presented in Table~\ref{tab2} should by no means be interpreted as predictive. Rather, these estimates provide an estimate for the potential size of the new markets created by the current transition in the steel sector, highlighting both the challenges of creating new supply networks for sustainable steelmaking and the business opportunities that this presents. A more thorough quantitative understanding of how this potential might be realized under different scenarios, requires a comprehensive modelling approach that also includes such supply- and demand-side factors.

There are several limitations to our work. Our trade network analysis relies on the categorisation of products into harmonised system codes. The scrap-related categories may also include types of ferrous waste that are not useful as input for EAFs. The data also does not include information on 'domestic' production of scrap, we only consider cross-border flows. The analysis at company level is based on a business intelligence database, which is one of the most comprehensive such datasets in the world, but does not contain a complete register of companies and might vary in coverage across countries. Note that we focus the modelling analysis on European countries, where the coverage of our data source is known to be more homogeneous \cite{/content/paper/c7bdaa03-en}.

Our search strategy may also have missed companies with relevant activities, especially for companies without a textual description of their main products and services. In terms of statistical modelling, we rely on extrapolation from phenomenological observations. To better understand the role of scaling effects and non-linear constraints (e.g. availability of critical inputs such as scrap sources, labour and renewable energy), the development of custom-built mechanistic firm-level simulation models is required, which is beyond the scope of this work.

In conclusion, our work shows that the green transition in the EU steel industry presents challenges and opportunities. In addition to energy availability, the transition requires a significant increase in access to scrap, while current trends point to a declining imports in European countries. China is playing a more important role in steel production, while it is decoupling from the rest of the world in terms of trade flows. The increased demand for scrap in Europe also creates a new market for companies, potentially creating around 30,000 jobs and companies with a combined turnover of around USD 35 billion.

\backmatter




















\bibliography{sn-bibliography}


\begin{thebibliography}{27}
\ifx \bisbn   \undefined \def \bisbn  #1{ISBN #1}\fi
\ifx \binits  \undefined \def \binits#1{#1}\fi
\ifx \bauthor  \undefined \def \bauthor#1{#1}\fi
\ifx \batitle  \undefined \def \batitle#1{#1}\fi
\ifx \bjtitle  \undefined \def \bjtitle#1{#1}\fi
\ifx \bvolume  \undefined \def \bvolume#1{\textbf{#1}}\fi
\ifx \byear  \undefined \def \byear#1{#1}\fi
\ifx \bissue  \undefined \def \bissue#1{#1}\fi
\ifx \bfpage  \undefined \def \bfpage#1{#1}\fi
\ifx \blpage  \undefined \def \blpage #1{#1}\fi
\ifx \burl  \undefined \def \burl#1{\textsf{#1}}\fi
\ifx \doiurl  \undefined \def \doiurl#1{\url{https://doi.org/#1}}\fi
\ifx \betal  \undefined \def \betal{\textit{et al.}}\fi
\ifx \binstitute  \undefined \def \binstitute#1{#1}\fi
\ifx \binstitutionaled  \undefined \def \binstitutionaled#1{#1}\fi
\ifx \bctitle  \undefined \def \bctitle#1{#1}\fi
\ifx \beditor  \undefined \def \beditor#1{#1}\fi
\ifx \bpublisher  \undefined \def \bpublisher#1{#1}\fi
\ifx \bbtitle  \undefined \def \bbtitle#1{#1}\fi
\ifx \bedition  \undefined \def \bedition#1{#1}\fi
\ifx \bseriesno  \undefined \def \bseriesno#1{#1}\fi
\ifx \blocation  \undefined \def \blocation#1{#1}\fi
\ifx \bsertitle  \undefined \def \bsertitle#1{#1}\fi
\ifx \bsnm \undefined \def \bsnm#1{#1}\fi
\ifx \bsuffix \undefined \def \bsuffix#1{#1}\fi
\ifx \bparticle \undefined \def \bparticle#1{#1}\fi
\ifx \barticle \undefined \def \barticle#1{#1}\fi
\bibcommenthead
\ifx \bconfdate \undefined \def \bconfdate #1{#1}\fi
\ifx \botherref \undefined \def \botherref #1{#1}\fi
\ifx \url \undefined \def \url#1{\textsf{#1}}\fi
\ifx \bchapter \undefined \def \bchapter#1{#1}\fi
\ifx \bbook \undefined \def \bbook#1{#1}\fi
\ifx \bcomment \undefined \def \bcomment#1{#1}\fi
\ifx \oauthor \undefined \def \oauthor#1{#1}\fi
\ifx \citeauthoryear \undefined \def \citeauthoryear#1{#1}\fi
\ifx \endbibitem  \undefined \def \endbibitem {}\fi
\ifx \bconflocation  \undefined \def \bconflocation#1{#1}\fi
\ifx \arxivurl  \undefined \def \arxivurl#1{\textsf{#1}}\fi
\csname PreBibitemsHook\endcsname

\bibitem[\protect\citeauthoryear{Somers et~al.}{2022}]{somers2022technologies}
\begin{bbook}
\bauthor{\bsnm{Somers}, \binits{J.}}, \betal:
\bbtitle{Technologies to Decarbonise the EU Steel Industry}.
\bpublisher{Publications Office of the European Union},
\blocation{Luxembourg}
(\byear{2022})
\end{bbook}
\endbibitem

\bibitem[\protect\citeauthoryear{voestalpine}{2023}]{voestcr}
\begin{botherref}
\oauthor{\bsnm{voestalpine}}:
Corporate Responsibility Report 2021/22,
Linz, Austria
(2023)
\end{botherref}
\endbibitem

\bibitem[\protect\citeauthoryear{Erhart and Erhart}{2023}]{erhart2023environmental}
\begin{barticle}
\bauthor{\bsnm{Erhart}, \binits{S.}},
\bauthor{\bsnm{Erhart}, \binits{K.}}:
\batitle{Environmental ranking of european industrial facilities by toxicity and global warming potentials}.
\bjtitle{Scientific Reports}
\bvolume{13}(\bissue{1}),
\bfpage{1772}
(\byear{2023})
\end{barticle}
\endbibitem

\bibitem[\protect\citeauthoryear{}{}]{EuroferProjects}
\begin{botherref}
Low-CO2 emissions projects in the EU steel industry.
\url{https://www.eurofer.eu/issues/climate-and-energy/maps-of-key-low-carbon-steel-projects}.
Accessed: 2024-03-12
\end{botherref}
\endbibitem

\bibitem[\protect\citeauthoryear{Vogl et~al.}{2018}]{vogl2018assessment}
\begin{barticle}
\bauthor{\bsnm{Vogl}, \binits{V.}},
\bauthor{\bsnm{{\AA}hman}, \binits{M.}},
\bauthor{\bsnm{Nilsson}, \binits{L.J.}}:
\batitle{Assessment of hydrogen direct reduction for fossil-free steelmaking}.
\bjtitle{Journal of cleaner production}
\bvolume{203},
\bfpage{736}--\blpage{745}
(\byear{2018})
\end{barticle}
\endbibitem

\bibitem[\protect\citeauthoryear{Swalec and Grigsby-Schulte}{2023}]{swalec2023pedal}
\begin{bbook}
\bauthor{\bsnm{Swalec}, \binits{C.}},
\bauthor{\bsnm{Grigsby-Schulte}, \binits{A.}}:
\bbtitle{Global Energy Monitor 2023. Pedal to the Metal.}
\bpublisher{Global Energy Monitor},
\blocation{Covina, CA}
(\byear{2023})
\end{bbook}
\endbibitem

\bibitem[\protect\citeauthoryear{Pauliuk et~al.}{2013}]{pauliuk2013steel}
\begin{barticle}
\bauthor{\bsnm{Pauliuk}, \binits{S.}},
\bauthor{\bsnm{Milford}, \binits{R.L.}},
\bauthor{\bsnm{Muller}, \binits{D.B.}},
\bauthor{\bsnm{Allwood}, \binits{J.M.}}:
\batitle{The steel scrap age}.
\bjtitle{Environmental science \& technology}
\bvolume{47}(\bissue{7}),
\bfpage{3448}--\blpage{3454}
(\byear{2013})
\end{barticle}
\endbibitem

\bibitem[\protect\citeauthoryear{Kim and Sohn}{2022}]{kim2022critical}
\begin{barticle}
\bauthor{\bsnm{Kim}, \binits{W.}},
\bauthor{\bsnm{Sohn}, \binits{I.}}:
\batitle{Critical challenges facing low carbon steelmaking technology using hydrogen direct reduced iron}.
\bjtitle{Joule}
\bvolume{6}(\bissue{10}),
\bfpage{2228}--\blpage{2232}
(\byear{2022})
\end{barticle}
\endbibitem

\bibitem[\protect\citeauthoryear{Manana et~al.}{2021}]{manana2021increase}
\begin{barticle}
\bauthor{\bsnm{Manana}, \binits{M.}},
\bauthor{\bsnm{Zobaa}, \binits{A.}},
\bauthor{\bsnm{Vaccaro}, \binits{A.}},
\bauthor{\bsnm{Arroyo}, \binits{A.}},
\bauthor{\bsnm{Martinez}, \binits{R.}},
\bauthor{\bsnm{Castro}, \binits{P.}},
\bauthor{\bsnm{Laso}, \binits{A.}},
\bauthor{\bsnm{Bustamante}, \binits{S.}}:
\batitle{Increase of capacity in electric arc-furnace steel mill factories by means of a demand-side management strategy and ampacity techniques}.
\bjtitle{International Journal of Electrical Power \& Energy Systems}
\bvolume{124},
\bfpage{106337}
(\byear{2021})
\end{barticle}
\endbibitem

\bibitem[\protect\citeauthoryear{Carlsson et~al.}{2020}]{carlsson2020modeling}
\begin{barticle}
\bauthor{\bsnm{Carlsson}, \binits{L.S.}},
\bauthor{\bsnm{Samuelsson}, \binits{P.B.}},
\bauthor{\bsnm{J{\"o}nsson}, \binits{P.G.}}:
\batitle{Modeling the effect of scrap on the electrical energy consumption of an electric arc furnace}.
\bjtitle{Processes}
\bvolume{8}(\bissue{9}),
\bfpage{1044}
(\byear{2020})
\end{barticle}
\endbibitem

\bibitem[\protect\citeauthoryear{Toma{\v{z}}i{\v{c}} et~al.}{2022}]{tomavzivc2022data}
\begin{barticle}
\bauthor{\bsnm{Toma{\v{z}}i{\v{c}}}, \binits{S.}},
\bauthor{\bsnm{Andonovski}, \binits{G.}},
\bauthor{\bsnm{{\v{S}}krjanc}, \binits{I.}},
\bauthor{\bsnm{Logar}, \binits{V.}}:
\batitle{Data-driven modelling and optimization of energy consumption in eaf}.
\bjtitle{Metals}
\bvolume{12}(\bissue{5}),
\bfpage{816}
(\byear{2022})
\end{barticle}
\endbibitem

\bibitem[\protect\citeauthoryear{{\L}ukasik and Olczykowski}{2020}]{lukasik2020estimating}
\begin{barticle}
\bauthor{\bsnm{{\L}ukasik}, \binits{Z.}},
\bauthor{\bsnm{Olczykowski}, \binits{Z.}}:
\batitle{Estimating the impact of arc furnaces on the quality of power in supply systems}.
\bjtitle{Energies}
\bvolume{13}(\bissue{6}),
\bfpage{1462}
(\byear{2020})
\end{barticle}
\endbibitem

\bibitem[\protect\citeauthoryear{EUROFER}{2019}]{roadmap2019pathways}
\begin{bbook}
\bauthor{\bsnm{EUROFER}}:
\bbtitle{Pathways to a CO2-Neutral European Steel Industry.}
\bpublisher{EUROFER},
\blocation{Brussels, Belgium}
(\byear{2019})
\end{bbook}
\endbibitem

\bibitem[\protect\citeauthoryear{Toktarova et~al.}{2022}]{toktarova2022interaction}
\begin{barticle}
\bauthor{\bsnm{Toktarova}, \binits{A.}},
\bauthor{\bsnm{Walter}, \binits{V.}},
\bauthor{\bsnm{G{\"o}ransson}, \binits{L.}},
\bauthor{\bsnm{Johnsson}, \binits{F.}}:
\batitle{Interaction between electrified steel production and the north european electricity system}.
\bjtitle{Applied Energy}
\bvolume{310},
\bfpage{118584}
(\byear{2022})
\end{barticle}
\endbibitem

\bibitem[\protect\citeauthoryear{Gielen et~al.}{2020}]{gielen2020renewables}
\begin{barticle}
\bauthor{\bsnm{Gielen}, \binits{D.}},
\bauthor{\bsnm{Saygin}, \binits{D.}},
\bauthor{\bsnm{Taibi}, \binits{E.}},
\bauthor{\bsnm{Birat}, \binits{J.-P.}}:
\batitle{Renewables-based decarbonization and relocation of iron and steel making: A case study}.
\bjtitle{Journal of Industrial Ecology}
\bvolume{24}(\bissue{5}),
\bfpage{1113}--\blpage{1125}
(\byear{2020})
\end{barticle}
\endbibitem

\bibitem[\protect\citeauthoryear{Devlin et~al.}{2023}]{devlin2023global}
\begin{barticle}
\bauthor{\bsnm{Devlin}, \binits{A.}},
\bauthor{\bsnm{Kossen}, \binits{J.}},
\bauthor{\bsnm{Goldie-Jones}, \binits{H.}},
\bauthor{\bsnm{Yang}, \binits{A.}}:
\batitle{Global green hydrogen-based steel opportunities surrounding high quality renewable energy and iron ore deposits}.
\bjtitle{Nature Communications}
\bvolume{14}(\bissue{1}),
\bfpage{2578}
(\byear{2023})
\end{barticle}
\endbibitem

\bibitem[\protect\citeauthoryear{Hu et~al.}{2020}]{hu2020characteristics}
\begin{barticle}
\bauthor{\bsnm{Hu}, \binits{X.}},
\bauthor{\bsnm{Wang}, \binits{C.}},
\bauthor{\bsnm{Lim}, \binits{M.K.}},
\bauthor{\bsnm{Koh}, \binits{S.L.}}:
\batitle{Characteristics and community evolution patterns of the international scrap metal trade}.
\bjtitle{Journal of Cleaner Production}
\bvolume{243},
\bfpage{118576}
(\byear{2020})
\end{barticle}
\endbibitem

\bibitem[\protect\citeauthoryear{Gaulier and Zignago}{2010}]{CEPII:2010-23}
\begin{botherref}
\oauthor{\bsnm{Gaulier}, \binits{G.}},
\oauthor{\bsnm{Zignago}, \binits{S.}}:
Baci: International trade database at the product-level. the 1994-2007 version.
Working Papers 2010-23,
CEPII
(2010).
\url{http://www.cepii.fr/CEPII/fr/publications/wp/abstract.asp?NoDoc=2726}
\end{botherref}
\endbibitem

\bibitem[\protect\citeauthoryear{Serrano et~al.}{2009}]{serrano2009extracting}
\begin{barticle}
\bauthor{\bsnm{Serrano}, \binits{M.{\'A}.}},
\bauthor{\bsnm{Bogun{\'a}}, \binits{M.}},
\bauthor{\bsnm{Vespignani}, \binits{A.}}:
\batitle{Extracting the multiscale backbone of complex weighted networks}.
\bjtitle{Proceedings of the national academy of sciences}
\bvolume{106}(\bissue{16}),
\bfpage{6483}--\blpage{6488}
(\byear{2009})
\end{barticle}
\endbibitem

\bibitem[\protect\citeauthoryear{van Dijk}{2023}]{van2023private}
\begin{botherref}
\oauthor{\bsnm{Dijk}, \binits{B.}}:
Private company information--Orbis.(nd). Bvd
(2023)
\end{botherref}
\endbibitem

\bibitem[\protect\citeauthoryear{Bajgar et~al.}{2020}]{/content/paper/c7bdaa03-en}
\begin{botherref}
\oauthor{\bsnm{Bajgar}, \binits{M.}},
\oauthor{\bsnm{Berlingieri}, \binits{G.}},
\oauthor{\bsnm{Calligaris}, \binits{S.}},
\oauthor{\bsnm{Criscuolo}, \binits{C.}},
\oauthor{\bsnm{Timmis}, \binits{J.}}:
Coverage and representativeness of orbis data
(2020)
\doiurl{10.1787/c7bdaa03-en}
\end{botherref}
\endbibitem

\bibitem[\protect\citeauthoryear{Foulds et~al.}{2013}]{foulds2013stochastic}
\begin{bchapter}
\bauthor{\bsnm{Foulds}, \binits{J.}},
\bauthor{\bsnm{Boyles}, \binits{L.}},
\bauthor{\bsnm{DuBois}, \binits{C.}},
\bauthor{\bsnm{Smyth}, \binits{P.}},
\bauthor{\bsnm{Welling}, \binits{M.}}:
\bctitle{Stochastic collapsed variational bayesian inference for latent dirichlet allocation}.
In: \bbtitle{Proceedings of the 19th ACM SIGKDD International Conference on Knowledge Discovery and Data Mining},
pp. \bfpage{446}--\blpage{454}
(\byear{2013})
\end{bchapter}
\endbibitem

\bibitem[\protect\citeauthoryear{Blei et~al.}{2003}]{blei2003latent}
\begin{barticle}
\bauthor{\bsnm{Blei}, \binits{D.M.}},
\bauthor{\bsnm{Ng}, \binits{A.Y.}},
\bauthor{\bsnm{Jordan}, \binits{M.I.}}:
\batitle{Latent dirichlet allocation}.
\bjtitle{Journal of machine Learning research}
\bvolume{3}(\bissue{Jan}),
\bfpage{993}--\blpage{1022}
(\byear{2003})
\end{barticle}
\endbibitem

\bibitem[\protect\citeauthoryear{Dodds et~al.}{2011}]{dodds2011temporal}
\begin{barticle}
\bauthor{\bsnm{Dodds}, \binits{P.S.}},
\bauthor{\bsnm{Harris}, \binits{K.D.}},
\bauthor{\bsnm{Kloumann}, \binits{I.M.}},
\bauthor{\bsnm{Bliss}, \binits{C.A.}},
\bauthor{\bsnm{Danforth}, \binits{C.M.}}:
\batitle{Temporal patterns of happiness and information in a global social network: Hedonometrics and twitter}.
\bjtitle{PloS one}
\bvolume{6}(\bissue{12}),
\bfpage{26752}
(\byear{2011})
\end{barticle}
\endbibitem

\bibitem[\protect\citeauthoryear{Wallach et~al.}{2009}]{wallach2009evaluation}
\begin{bchapter}
\bauthor{\bsnm{Wallach}, \binits{H.M.}},
\bauthor{\bsnm{Murray}, \binits{I.}},
\bauthor{\bsnm{Salakhutdinov}, \binits{R.}},
\bauthor{\bsnm{Mimno}, \binits{D.}}:
\bctitle{Evaluation methods for topic models}.
In: \bbtitle{Proceedings of the 26th Annual International Conference on Machine Learning},
pp. \bfpage{1105}--\blpage{1112}
(\byear{2009})
\end{bchapter}
\endbibitem

\bibitem[\protect\citeauthoryear{Ren et~al.}{2021}]{ren2021review}
\begin{barticle}
\bauthor{\bsnm{Ren}, \binits{L.}},
\bauthor{\bsnm{Zhou}, \binits{S.}},
\bauthor{\bsnm{Peng}, \binits{T.}},
\bauthor{\bsnm{Ou}, \binits{X.}}:
\batitle{A review of co2 emissions reduction technologies and low-carbon development in the iron and steel industry focusing on china}.
\bjtitle{Renewable and Sustainable Energy Reviews}
\bvolume{143},
\bfpage{110846}
(\byear{2021})
\end{barticle}
\endbibitem

\bibitem[\protect\citeauthoryear{}{}]{worldsteel}
\begin{botherref}
Scrap use in the steel industry.
\url{https://worldsteel.org/wp-content/uploads/Fact-sheet-on-scrap_2021.pdf}.
Accessed: 2024-03-12
\end{botherref}
\endbibitem

\end{thebibliography}

\end{document}